\documentclass[reprint,onecolumn,amsmath,amssymb,aps,
]{revtex4}

\usepackage{epsfig,bm,epsf,graphics}
\usepackage{graphicx}
\usepackage{dcolumn}
\usepackage{bm}
\usepackage{xcolor}

\begin{document}
\preprint{APS/123-QED}

\title{Phase Transition and Magneto-caloric Properties of Perovskites Pr$_{0.55}$Sr$_{0.45}$MnO$_{3}$: Modeling versus Experiments}

\author{Yethreb Essouda$^{1}$\footnote{yethreb.essouda@gmail.com},  Hung T. Diep$^{2}$\footnote{diep@cyu.fr, corresponding author,https://orcid.org/0000-0003-1852-3435}, Mohamed Ellouze$^1$\footnote{mohamed.ellouze@fss.usf.tn, https://orcid.org/my-orcid?orcid=0000-0002-5662-8572 }}
\affiliation{%
$^1$Sfax University, Faculty of Sciences of Sfax, LM2EM, B.P. 1171, 3000, Sfax, Tunisia.\\
$^2$  Laboratoire de Physique Th\'eorique et Mod\'elisation,
CY Cergy Paris Universit\'e, CNRS, UMR 8089\\
2, Avenue Adolphe Chauvin, 95302 Cergy-Pontoise Cedex, France.
}



\date{\today}


\begin{abstract}

Experimental data obtained with the perovskite compounds  Pr$_{0.55}$Sr$_{0.45}$MnO$_{3}$ show that the magnetization decreases with increasing temperature $T$ and undergoes a very sharp phase transition  to the paramagnetic phase.  The sharp transition in a system with a strong disorder is very rare, if not non-existent, in the theory of phase transition in systems of short-range  pairwise exchange interactions. 
To understand  this remarkable property, we introduce a model including a  multispin (cluster-like) interaction  between Mn ions, in addition to the usual pairwise exchange terms between these ions and the Mn-Pr interactions.  We carry out Monte Carlo (MC) simulations using this model. The crystal is a body-centered tetragonal lattice where the corner sites are occupied by Mn ions and the center sites by Pr or Sr ions.  Due to the doping, Mn$^{4+}$ with $S=3/2$ has the concentration of Pr$^{3+}$ ($S=1$) and Mn$^{3+}$ with $S=2$ has the Sr concentration.  After attempts with different spin models and various Hamiltonians, we find that the many-state Ising spin model reproduces most of the experimental results. For the Hamiltonian, we find that pairwise interactions alone between ions cannot reproduce the sharp transition and the magnetization  below $T_C$.  We have to include a multispin interaction as said above.  
We fit the MC results with experimental data, and we estimate values of various exchange interactions in the system.  These values are found to be in the range of those found in perovskite manganite compounts. We also study the applied-field effect on the magnetization in the temperature region below and above the transition temperature $T_C$.  We calculate the magnetic entropy change $|\Delta S_m|$ and the  Relative Cooling Power,  for magnetic field from 1 to 3 Tesla. Our simulation results  are in good agreement with experiments.  The role of the multispin interaction is analyzed and discussed.    
\vspace{0.5cm}
\begin{description}
\item[PACS numbers: 5.10.Ln;64.30.-t;75.50.Cc]
\item[Keywords: Praseodymium strontium manganite (PSMO); Phase transition; Magnetic entropy change;]
\item[Relative Cooling Power; Monte Carlo simulation]
\end{description}
\end{abstract}

\maketitle


\section{INTRODUCTION}

Interest in seeking broad clean air protections and environmentally green technology to reduce production and consumption of hydrofluorocarbons (HFCs) has continued to grow over the last decades. Much ink has been spilled on the topic of ‘’innovation new technology for cooling ‘’ in view to replace refrigerants with high global warming potential (GWP). Among the most cooling technologies used today across the world, magnetic refrigeration can be considered as a successfully solution that may offer larger efficiencies than a traditional refrigeration technique owing its promissing characgteristics. This emergent technology could provide efficient heating and cooling, which accounts for more than half of the energy used in homes, and help phase out conventional gas-compression refrigeration [1], which use gases with high GWP as refrigerants. The magnetic refrigeration, based on magnetocaloric effect (MCE) would eliminate the risk of such gases escaping into the atmosphere by replacing them with heating or cooling of a magnetic material upon application or removal of magnetic field which makes it environmentally safe refrigerant with zero GWP [2]. Therefore, materials with magnetocaloric properties are suitable for use in refrigeration and air conditioning. Both a large, isothermal entropy change ($\Delta S_m$) induced by a variation of the external magnetic field [3] and adiabatic temperature change ($\Delta T$) determine the large MCE for a magnetic material. As noted earlier, the gadolinium (Gd) metal is mostly conspicuous magnetic refrigeration with a large MCE but has a drawback and restriction for its use in the actual application because  it has a high cost [4]. With the goal to innovate new substance with low price and large MCE, a lot of research has been done on magnetic materials that offered the best magnetic properties and cheaper than Gd. The rare-earth manganite perovskites with the general formula Ln$_{1-x}$A$_x$MnO$_3$ (Ln= trivalent rare earth, A=divalent alkaline earth) are prospect candidates for electronic and/or magnetic devices due to their large range of features such as colossal magnetoresistance (CMR) and MCE [5]. Recently, many studies on Pr-based manganites  have shown their striking properties including the charge ordering (CO) state, coexistence of ferromagnetic (FM) and antiferromagnetic (AF) phases, and metamagnetic transition [6-8]. Especially, the perovskite Pr$_{1-x}$Sr$_x$MnO$_3$ is a great promise member in manganites family with intermediate one electron bandwidth [9].

 More recently, Fan et al [10] have investigated experimental magnetic properties of perovskite manganite Pr$_{0.55}$Sr$_{0.45}$MnO$_3$ by using traditional solid-state reaction method. They reported that doping 45\% of Sr in PrMnO$_3$ can lead to a sharp paramagnetic (PM) - FM phase transition at 291K, and indicate that the present material possesses a large Relative Cooling Power (RCP) and a wide temperature range. These features raise Pr$_{0.55}$Sr$_{0.45}$MnO$_3$  to be a perfect candidate for magnetic refrigeration in the room temperature range. Motivated by the experimental progress of the MCE towards the doping of Pr with Sr, we use in this paper a modeling which contains various interaction terms between different kinds of magnetic ions in the system. As will be shown below, the "multi-state" Ising spin model with different spin amplitudes according to the kind of ions [Mn$^{3+}$: $S = 2$, Mn$^{4+}$:$S = 3/2$, Pr$^{3+}$: $S=1$ ] is the best to get an agreement with experimental data [10].   In addition to the  nearest-neighbor (NN) pairwise interactions, we have used also a multi-spin interaction between Mn ions. Using Monte Carlo (MC) simulations, we show that this term is  the origin of the sharp phase transition between FM and AF phases occurring at $T_C=291$ K [10].  Our model thus reproduces to a good agreement  experimental measurements  performed on this material: magnetization versus $T$ (temperature) and the magneto-caloric properties .  

The paper is organized as follows: in section II, we describe our  theoretical model and the MC method. Results are compared to experimental data in section III. Our concluding remarks are given in section IV.

\section{Model and Method}
\subsection{Experimental samples}
In this paper, we propose a theoretical model and carry out MC simulations based on that model. Our purpose is  to compare our MC  results with experimental data  obtained for Pr$_{0.55}$Sr$_{0.45}$MnO$_3$ [10].   Polycrystalline Pr$_{0.55}$Sr$_{0.45}$MnO$_3$  sample was prepared by traditional solid-state reaction method. The structure and phase purity of the prepared sample were verified by powder X-ray diffraction with Cu Ka radiation at room temperature.

We note that this  compound of the same concentration has also been prepared by another method:the polycrystalline sample of Pr$_{0.55}$Sr$_{0.45}$MnO$_3$  has been
synthesized using nitrate route [11]. The authors have described the sample preparation as follows: "The stoichiometric ratio of powder Pr$_6$O$_{11}$, SrCo$_3$ and MnO$_2$ were grounded and calcined several times between 800 $^{\circ}$C and 1200 $^{\circ}$C for 24 h with intermediate grindings. The powders thus obtained were pressed into pellet form at 10 MPa pressure and finally sintered at 1400 $^{\circ}$C for 30 h. The structure and phase formation were confirmed by powder X-ray diffraction (XRD) pattern recorded at room temperature and was refined by Rietveld method using FullProf Software program. The XRD pattern of the sample crystallized in an orthorhombic structure with Pnma space group comprising lattice parameters $a$ = 5.423(6) Å, $b$ = 7.660(7) Å and $c$ = 5.478(7) Å with cell volume $V$ = 227.632(1) Å$^3$.

The two above samples with the same formula do not however yield the  same experimental data:  in Ref. [10] there is only a transition from the paramagnetic (PM) phase to the ferromagnetic (FM) phase at $T_C=291 ^{\circ}$C, and this phase persists to near $T=0$. On the other hand, Ref. [11] finds that the PM-FM transition occurs at $T_C=281^{\circ}$C and the FM phase is transformed into the antiferromagnetic (AFM) phase at $T_C=181^{\circ}$C and this latter phase goes down to near $T=0$.  Does the way of sample preparation yield this difference?  We believe that this difference comes from domains and dislocations which are certainly numerous in the sample of Ref. [11] causing two phase transitions. There may be two types of domain, antiferromagnetic and ferromagnetic ones, in the sample of Ref. [11], and each type of domain  may have its own phase transition. To simplify the modeling, we choose to consider the experimental data of Ref.  [10] in this paper.

\subsection{Model}

\begin{figure}[ht]
\centering
\includegraphics[width=8cm,angle=0]{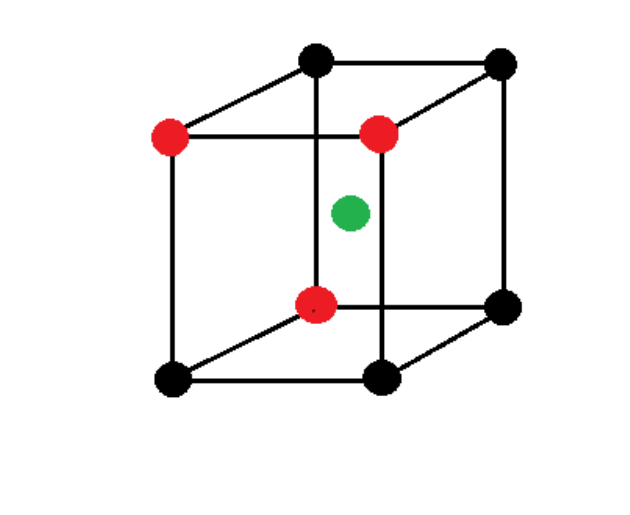}
\caption{A random ion distribution on the bct lattice: black, red and green circles represent Mn$^{3+}$, Mn$^{4+}$ and Pr$^{3+}$, respectively. }
\label{fig1}
\end{figure}

For the modeling, we use the body-cencered tetragonal (bct) lattice where the $c$-axis has the longuest lattice constant ( 7.66 Å) and the other two axes have approximately the same lattice constant ( 5.45 Å). The corner sites of the bct lattice are occupied either by Mn$^{3+} (S=2)$ or Mn$^{4+}(S=3/2)$ with their respective concentrations 0.55 and 0.45.  The centered sites are occupied by either Pr$^{3+}$ ions ($S=1$) or by non magnetic ions Sr with their respective concentrations 0.55 and 0.45.   We generate a random mixing of  Mn ions on the corner sites and Pr and Sr on the centered sites, respecting their respective concentrations.  This is a strongly disordered system.  An example of ion distribution in a bct cell is shown in Fig. \ref{fig1}.

For  the spin model, we shall use the  "multi-state" Ising model for all types of ion with different lengths $S=2$ for  Mn$^{3+}$, $S=3/2$ for Mn$^{4+}$, and $S=1$ for Pr$^{3+}$. Note that the two outer electrons of the Pr ion occupy the orbitals 4f2, so its spin is 1 (Hund's rule), and $L=3+2=5$. So, for less than half filled shell , $J=|L-S|=4$. However, as we are interested in spin-spin interactions of the compound which are responsible for the magnetic transition (not $J-J$ interaction), the orbitals $L$ are not taken into account. As for Mn$^{3+}$ and  Mn$^{4+}$,    Mn$^{3+}$ and Mn$^{4+}$ have spins 2 and 3/2, respectively by looking at the electronic configurations of those ions : the outer electrons of these ions occupy  3d$^4$  and 3d$^3$, respectively. By the Hund’s rule, 3d$^4$ of  Mn$^{3}$ yields $S=4\times 1/2=2$, and 3d$^3$ of Mn$^{4}$ yields $S=3\times1/2=3/2$, with their respective concentrations 0.55 and 0.45.

 Unlike the conventional Ising model with two spin states $\pm S$, we shall use the multi-state Ising model: the spin $S$ has $2S+1$ states $-S,-S+1,...,S-1,S$.  This model is somewhat similar to a quantum spin model, except the fact that it does not have $x$ and $y$ components, so it does not have quantum fluctuations.  This multi-state Ising model  expresses a strong anisotropy along the $c$-axis,  in addition to the difference of state numbers of   Mn$^{3+}$,  Mn$^{4+}$ and  Pr$^{3+}$ . We will discuss about this choice in section \ref{discuss}.

Let us define the pairwise interactions between magnetic ions by

\begin{equation}\label{pairwise}
 {\cal H}_p = -\sum_{<i,j>}J_{ij}\mathbf S_{i}\cdot \mathbf S_{j}-\mu_0H\sum_{<i>}S_{i}
\end{equation}
where $J_{ij}$ is the interaction strength between the  spins $S_i$ and $S_j$ occupied the sites $i$ and $j$, $H$ denotes the applied magnetic field applied in the +$z$ ($c$-axis) direction. We suppose only the interactions between nearest neighbors (NN) in the three directions of the bct lattice. We denote these  interactions as follows:

\indent $J_{1}$: Interaction coupling of a Mn$^{3+}$ ion with a NN Mn$^{3+}$ ion,\\ 
\indent $J_{2}$: Interaction coupling of a Mn$^{3+}$ ion with a NN Mn$^{4+}$ ion,\\  
\indent $J_{3}$: Interaction coupling of a Mn$^{4+}$ ion with a NN Mn$^{4+}$ ion,\\  
\indent $J_{4}$: Interaction coupling of a Pr ion with a Mn$^{3+}$ ion,\\
\indent $J_{5}$: Interaction coupling of a Pr ion with a Mn$^{4+}$ ion\\
\indent $J_{6}$: Interaction coupling between two Pr ions on the adjacent bct units.\\

Note that due to the longer lattice constant in the $c$-axis, we apply  a reduction coefficient $C=0.5$ on the interaction between NN  ions lying on the $c$-axis. A slight variation of $C$ causes of course a small shift on the value of $T_C$ but no new physics is found.

\begin{figure}[ht]
\centering
\includegraphics[width=8cm,angle=0]{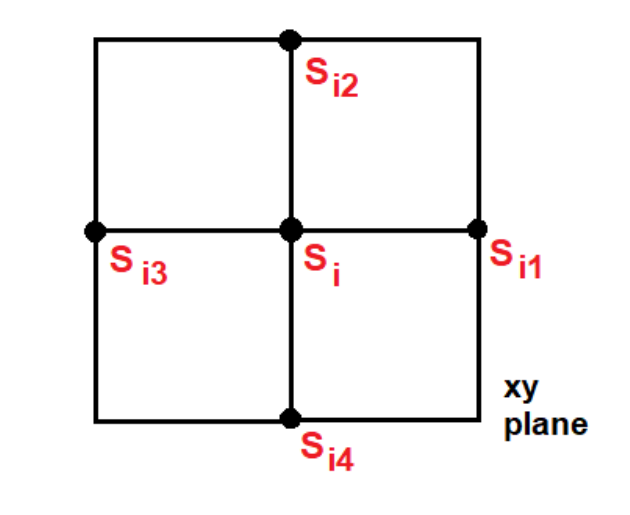}
\caption{ Spin $S_i$,of the Mn ion at the site $i$ interacts simultaneously with four Mn spins  $S_{i1}$, $S_{i2}$, $S_{i3}$ and $S_{i4}$ in the $xy$ plane. The multi-spin interaction is given by Eq. (\ref{multi}). See text for comments.  }
\label{fig2}
\end{figure}

The transition experimentally observed for our compound has a sharp first-order-like transition at $T_C$.  Our compound has a very strong disorder with a 0.55-0.45 mixing of Mn ions and with 0.45 percents of centered sites without spins. Such a strongly disordered system cannnot have a sharp phase transition as we see in spin glasses and amorphous magnetic materials.  For the modeling, we have tried the Heisenberg model, the continuous Ising model, ... with various pairwise interactions, but we did not find the magnetization curve with a first-order-like transition experimentally observed. The idea to introduce the multispin interaction to reproduce the experimental observations, as we will see later, was successful.   After several trial forms, the following cluster-like interaction between NN spins in the $xy$ plane, in addition to the above pairwise interactions, is written by

\begin{equation}\label{multi}
 {\cal H}_m = - K\sum_{i} S_{i} S_{i1} S_{i2} S_{i3} S_{i4}
\end{equation}
where $K$ is the interaction strength and the sum runs over all Mn sites and the spins $ S_{i1}, \ S_{i2},\  S_{i3}$ and $ S_{i4}$ are the NN of the spins $S_i$ on the $xy$ plane (see Fig. \ref{fig2}). 

The similar  model has been used with success to describe the magnetization plateau at low $T$ and the magnetic entropy change under an applied magnetic field in agrrement witrh experiments performed for the compound with another doping concentration, namely Pr$_{0.9}$Sr$_{0.1}$MnO$_3$ [12]. However, for the strong doping as in Pr$_{0.55}$Sr$_{0.45}$MnO$_3$ considered in this paper, we have to refine the spin model by using the multi-state Ising model, and to introduce the so-called Creutz over-relaxation in the MC flipping procedure as descibed below. In doing so, we obtain a very good agreement with experimental data. Note that magnetic properties strongly change with doping concentration. A simple model cannot interpret results of different doping concentrations. A model refinement is necessary as shown in this paper.  

 Let us  discuss about the muti-spin interaction.  The pairwise Ising model can  be considered as the one-component limit of the Hesenberg model. The pairwise Heisenberg model involves only two interacting spins because of the initial assumption of the overlap between the spin-dependent wave functions of only two neighboring atoms.  With the Hartree-Fock approximation, we can derive the Heisenberg model commonly used in the literature (see for example chapter 3 "Magnetic Exchange Interaction" of  Ref. [13]). In materials, however the interaction between one spin  with its neighbors is simultaneous, but the demonstration for the multi-spin interaction as in the case of two-body Heisenberg model is at present impossible, though  a fourth-order pertubation expansion for the Heisenberg model gives rise to a term of 3-spin  and a term of 4-spin interactions [14].  But for the Ising model, the demonstration of multispin interactions  by exact methods in statistical physics has been done. The reader is referred to Ref. [15] for the references on the demonstration of various multi-spin interactions. There has been an increasing number of papers studying the effect of the multi-spin interaction in various systems [15-17]. This domain is rather recent, very little is known on the properties stemming from such multi-spin interaction. In this regard, our multi-spin interaction (\ref{multi}) is original in materials science. We have shown  in [12] that it allows to interpret experimental data such as magnetization plateau below $T_C$. 

Let us give some  remarks on the multi-spin Hamiltonian (\ref{multi}).  We take a single plaquette which is shown in Fig. \ref{fig2}: we see that if two or four of the 5 spins of (\ref{multi}) change their signs, the energy does not vary. There is a large number of such choices of those flipping spins in a plaquette. However, when the plaquette is embedded in the lattice, such an invariance of the energy is verified only when the energy earn and the energy loss of such multi-spin flipping is conpensated, taking into account the neighboring spins of the plaquette. We conjecture that there is more chance for the energy invariance to occur, specially when the system is highly excited, near $T_C$ for example. This will overcome the so-called "critical slowing-down" near $T_C$ known in the theory of phase transitions and makes the transition dynamically easier.  We will see later that this term makes the transition sharper which is in agreement with experimental data of Ref. [10].  We will comment and compare our results with [10,11]. 

\subsection{Monte Carlo Method}

For MC simulations, we used   samples of dimension $N=L\times L\times L$, where $L$ is the linear system dimension, i. e. the  number of bct cells in each of the $x$, $y$ and $z$ directions. We used the periodic boundary conditions in all directions to reduce the surface effects. We have performed simulations with lattice sizes from $12^3$ to $30^3$ lattice cells to check finite-size effects.  The results shown below are those of $30^3$ lattice size. Note that no significant finite-size effect is seen from  $20^3$.

The first step is to generate a configuration of magnetic ions respecting their concentrations.  No configuration with a deviation more than 1\% from the compound formula is accepted.  We next  thermalize the spin configuration, starting from a random spin configuration as the initial condition, at a given temperature $T$.  For the MC spin update, we use the Metropolis algorithm [18,19] modified by an over-relaxation procedure [20,21].  As said above, the spins of the model is multi-state: for example the spin of Mn$^{3+}$ is $S=2$, for the spin update we have to take at random one state among $(2S+1)$ states: -2, -1, 0, +1, +2, and to use the updating algorithm as follows: 

(i) At the time $t$, we take a spin $i$ and calculate the field $H_i$ acting on it from all neighboring spins, we calculate the energy of that spin in its current state $E_{old}=-S_iH_i$. 

(ii) Now, we take  at random a new spin state, called $S_i'$  among $(2S+1)$ states and calculate the new engergy $E_{new}=-S_i'H_i$.  Note that $H_i$ is the same. If $E_{new}\leq  E_{old}$ then the new state $S_i'$ is accepted. If $E_{new} >  E_{old}$ the new state is accepted only if $R < \exp [-\beta(E_{new}- E_{old}]$ where $R$ is a random number between 0 and 1, and $\beta=1/(k_BT)$. 

(iii) Given the fact that our system strongly fluctuates, we introduce the over-relaxing procedure due to Creutz [20,21]. The principle is very simple: we take the advantage of the fact that the acting field $H_i$ on spin $S_i$ does not change, we repeat  
the second step (ii) several times (usually 5 to 10 times), expecting that the spin under consideration $S_i$ has more chance to get an optimal orientation (this "Creutz" over-relaxation [20,21] has been shown in many works to  be very efficient for systems with strong fluctuations such as frustrated spin systems [19,22]).  We go to another spin and repeat steps (i)-(iii) until all spins are visited. We say we complete a MC step per spin.

(iv) When equilibrium is reached after a large number of  MC per spin,  we determine thermodynamic properties by taking thermal averages of various physical quantities defined below.

We have calculated
the internal energy per spin $ E$, the specific heat $C_V$, the total magnetization $M$, the magnetic susceptibility $\chi$, the spin length of each sublattice, as functions of temperature $T$ and magnetic field $H$.
The MC run time for equilibrating is about $10^5$ MC steps per spin. The averaging is taken, after equilibrating, over $10^5$ MC steps. A large number of runs with different starting ion configurations have been carried out to check the reproductivity of results shown below.

 The statistical averages of the $z$ (or $c$) spin component of Mn$^{3+}$ and Mn$^{4+}$ and Pr$^{3+}$ ($\langle S_1 \rangle$, $\langle S_2\rangle$ and $\langle S_3 \rangle$, respectively)  are defined by
\begin{equation}
\langle S_{\ell} \rangle=\frac{1}{N_{\ell}}\langle\sum_{i\in \ell}S_i\rangle \label{msub}
\end{equation}
where $\langle...\rangle$ indicates the statistical time average and the sum is taken over Mn$^{3+}$ ($\ell=1$) or Mn$^{4+}$ ($\ell=2$) or Pr$^{3+}$ ($\ell=3$), with $N_{\ell}$ being the number of spins of each kind. 

The total magnetization $M$ is defined by
\begin{equation}
\langle M\rangle =g\mu_B (\langle S_{1}\rangle +\langle S_{2}\rangle+\langle S_{3}\rangle) \label{mtot}
\end{equation}
where $g$ and $\mu_B$ are the "effective" gyromagnetic factor and the Bohr magneton. Note that in experiments, the magnetizations of Mn$^{3+}$, Mn$^{4+}$ and Pr$^{3+}$ were not measured separately, therefore the gyromagnetic factor $g$ which relates the total spin to $M$ should be understood as an "effective $g$".

The average internal energy $E$ per spin, the specific heat $C_V$ per spin and the susceptibility $CHI$ per spin are defined by
\begin{eqnarray}
\langle E(T) \rangle &=&\frac{1}{N}\langle ({\cal H}_p+{\cal H}_m)\rangle \\
C_V(T)&=&\frac{1}{k_BT^2}[\langle E^2\rangle - \langle E\rangle ^2]\label{CV}\\
\chi(T)&=&\frac{1}{k_BT} [\langle M^2\rangle -\langle M\rangle^2]\label{Chi}
\end{eqnarray}

Note that since our purpose is not to calculate the critical exponents of the transition, we use just the Metropolis algorithm modified by introducing the Creutz over-relaxation as described above to determine thermodynamic characteristics of the phase transition. Other advanced MC methods such as the histogram method or the Wang-Landau technique are of course useful in complicated systems such as frustrated spin systems where the order of the phase transition cannot be determined otherwise (see implementation of these methods for example in [23,24,25]).

\section{Monte  Carlo Results - Comparison with Experiments}

After many tests on the values of interaction, we find the following set which gives the best agreement between MC results and experimental data from Ref. [10]:

\begin{eqnarray}
J_1&=&-0.1J,\  J_2= +0.54J,\  J_3=-0.1J,\  J_4= -0.07J, \ J_5= -0.07J, \ J_6= 0.30J \label{val1}\\
C&=&0.5  \label{val2}\\
K&=&0.62  \label{val3}
\end{eqnarray}
where $J$ is equal  to 1, namely $J$ is the MC energy unit. $C$ is the reduction coefficient applied to the interaction between ions on the $c$ axis.  Experimentalists believe that the interaction between Mn ions of the same kind is weakly antiferromagnetic while that between Mn$^{3+}$ and Mn$^{4+}$ is strongly ferromagnetic so that the overall ordering is ferromagnetic. That explains our choice of the interaction signs above. 

Note that the system is very sensitive to even very small variations of the above values.  A fine tuning was necessary.  In a word, $J_2$ competes with $K$ at low $T$: $J_2$ favors the ferromagnetic ordering while $K$ (multi-spin interaction) favors many states which destroy the ferromagnetic ordering. Near $T_C$, $K$ competes with the inter-sublattice interaction $J_4$-$J_6$ [see the discussion below Eq. (\ref{multi})].

 Using the above interaction values we obtained the MC transition temperature $T_C(MC)=0.7832$ in MC unit of $J=1$.  
In order to fit the MC transition temperature with the experimental value $T_C(exp)=291$ K [10], we have to multiply all the interaction values given above by $J=291/0.7832\simeq 371.55$.  This is possible because $T_C$ is proportional to the exchange interactions. 

\begin{figure}[ht]
\centering
\includegraphics[width=7cm,angle=0]{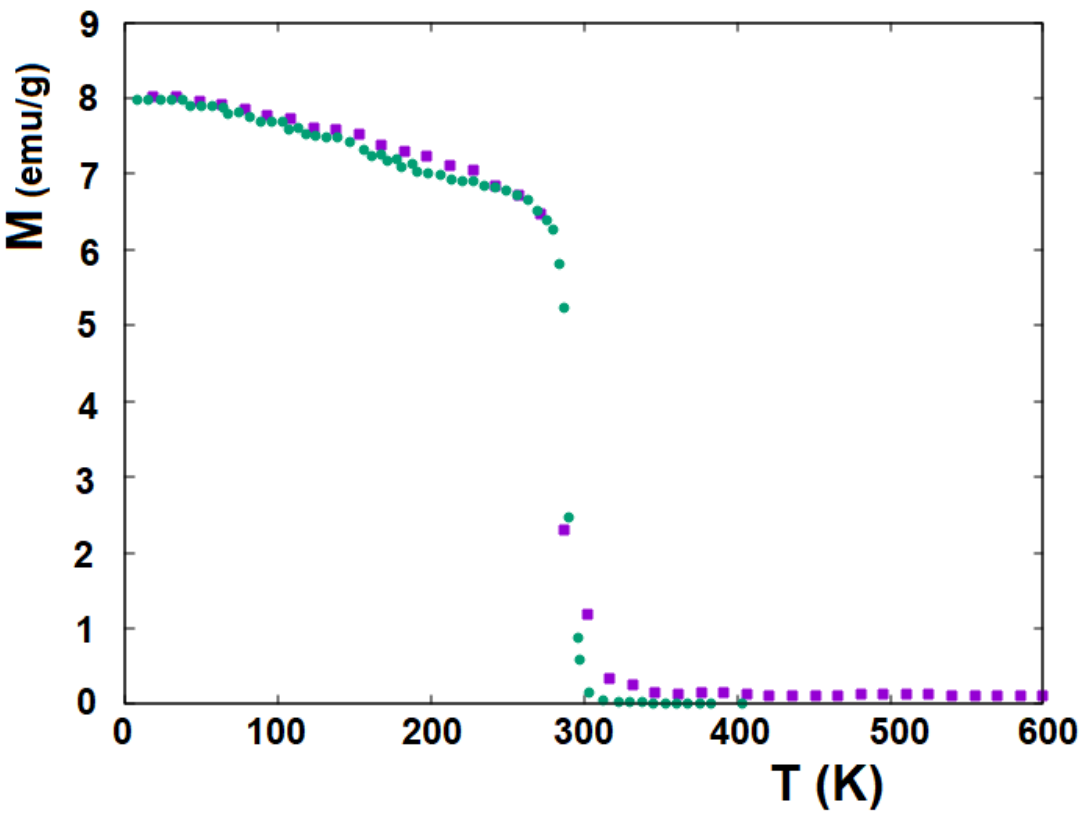}
\includegraphics[width=7cm,angle=0]{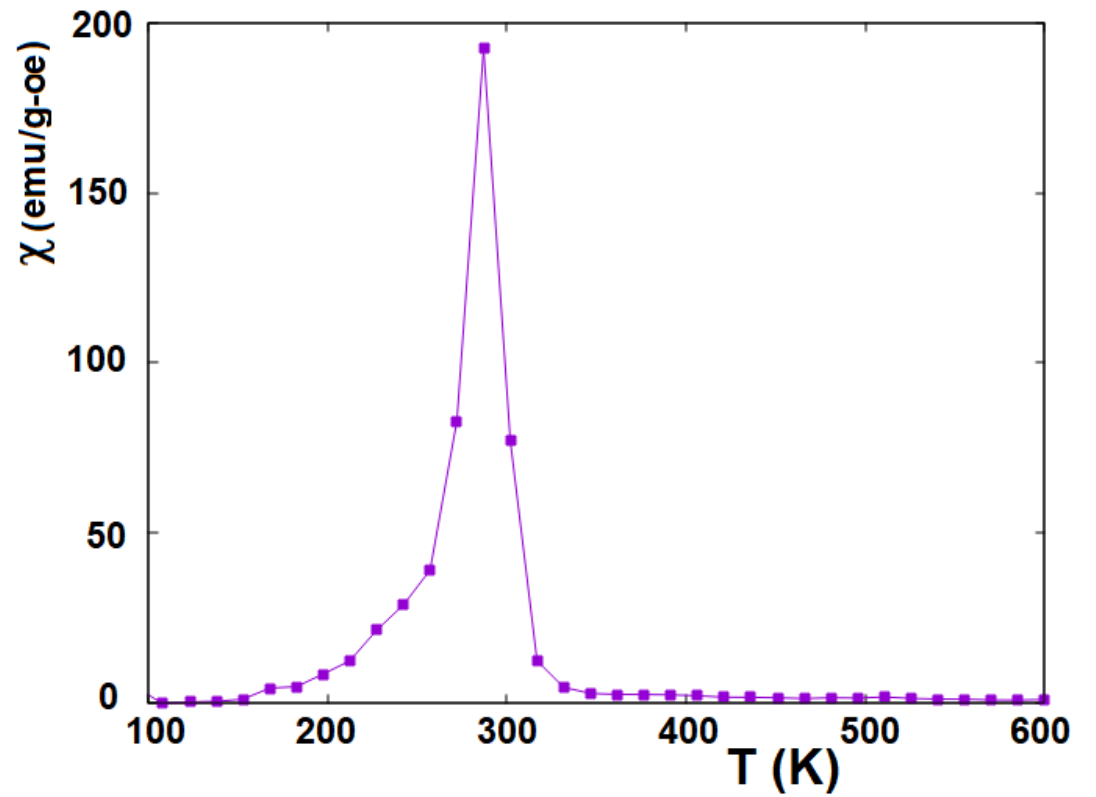}
\caption{(a) MC result (violet filled squares) and experimental magnetization (green void squares) taken from Ref. [10], versus  temperature $T$ in Kelvin, are shown for comparison; (b) MC result of susceptibility $\chi(T)$. Line is guide to the eye. See text for comments.  }
\label{fig3}
\end{figure}

We show in Fig. \ref{fig3}a the MC result of the total magnetization $M$ as a function of $T$ fitted  with the experimental $M$ taken from Ref. [10], using $J=371.55$. As seen,  except at an intermediate region of temperature, the MC results coincide with the  experimental curve at low $T$ and strikingly at $T_C$. The transition is very sharp, but it is not a first-order transition as discussed below. The total magnetic susceptibility $\chi(T)$ is shown in Fig. \ref{fig3}b. One observes here a very high peak of $\chi(T)$ at $T_C$.  
Since $\chi(T)$ is calculated by magnetization fluctuations using Eq. (\ref{Chi}) the huge  peak is the proof  that the magnetization very strongly fluctuates at $T_C$.

Let us note that with the above set of parameters, we  did not find the results of Ref. [11]. We will return to this point in section \ref{discuss}.

\begin{figure}[ht]
\centering
\includegraphics[width=8cm,angle=0]{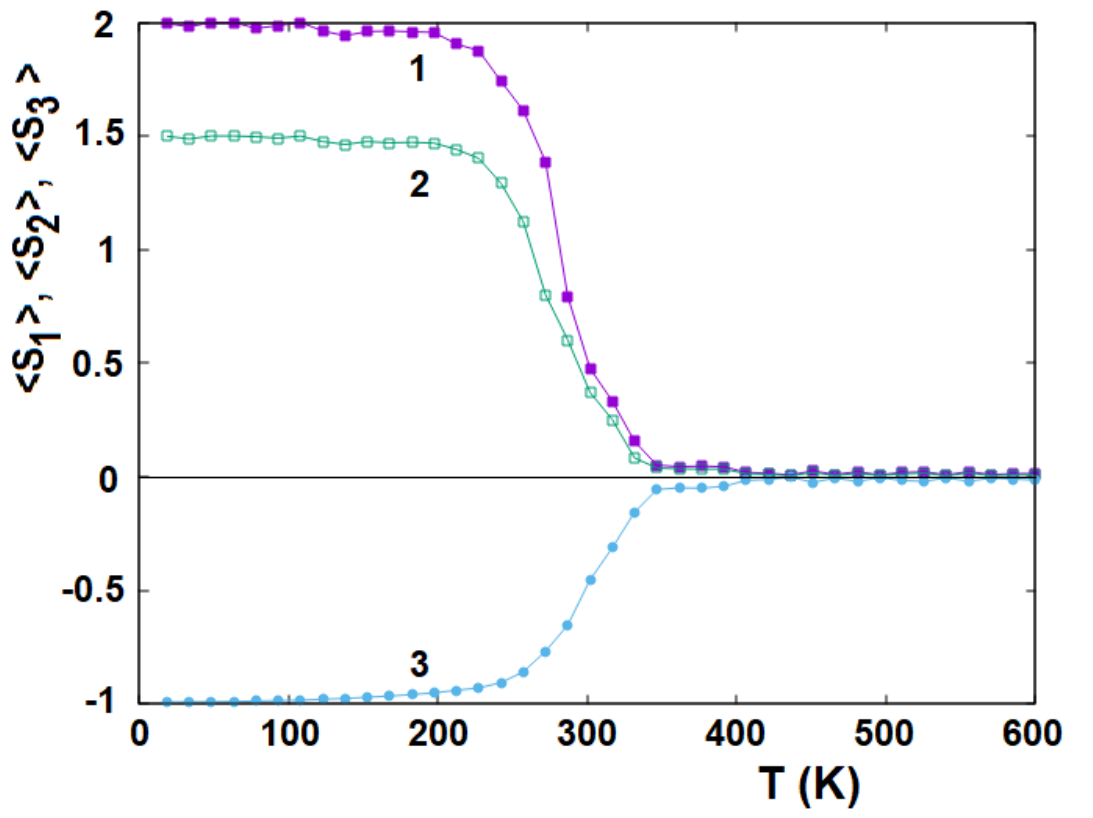}
\caption{ Sublattice  spin lengths $\langle S_1 \rangle$, $\langle S_2\rangle$ and $\langle S_3\rangle$   versus  temperature $T$ in unit of Kelvin. Curves 1, 2 and 3 correspond respectively to  $\langle S_1 \rangle$, $\langle S_2\rangle$ and $\langle S_3\rangle$ .  See text  for comments. }
\label{fig4}
\end{figure}

We show now in Fig. \ref{fig4} the spin lengths of three sublattice magnetic ions $\langle S (i) \rangle\ (i=1,2,3)$.   We see here that the Mn$^{3+}$ ($S(1)$) and  Mn$^{4+}$ ($S(2)$) have the same sign, namely they order ferromagnetically, while $S(3)$ (Pr ions) is ordered antiferromagnetically with the Mn ions.  These data are MC results. Experiments did not have access to these. Note the smooth variation of these quantities across  $T_C$. These components vanish at $T\simeq 350$ K as seen in  Fig. \ref{fig4}.   A question is naturally raised: given this smooth variation of spin lengths, how can the total magnetization shown  in Fig. \ref{fig3} undergo a vertical fall at $T_C$?  We have analyzed the MC data and found that at a temperature where the spin lengths are far from zero ($<S_3>\simeq -0.5$, $<S_1>\simeq 0.26$, $<S_2>\simeq 0.24$) the cancellation of $\langle S_1 \rangle+\langle S_2\rangle+\langle S_3\rangle$ occurs  causing the magnetization to fall vertically to zero at that $T$ which was defined as $T_C$ (291 K).  This is a beautiful effect of compensation of sublattice magnetizations.

\begin{figure}[ht]
\centering
\includegraphics[width=8cm,angle=0]{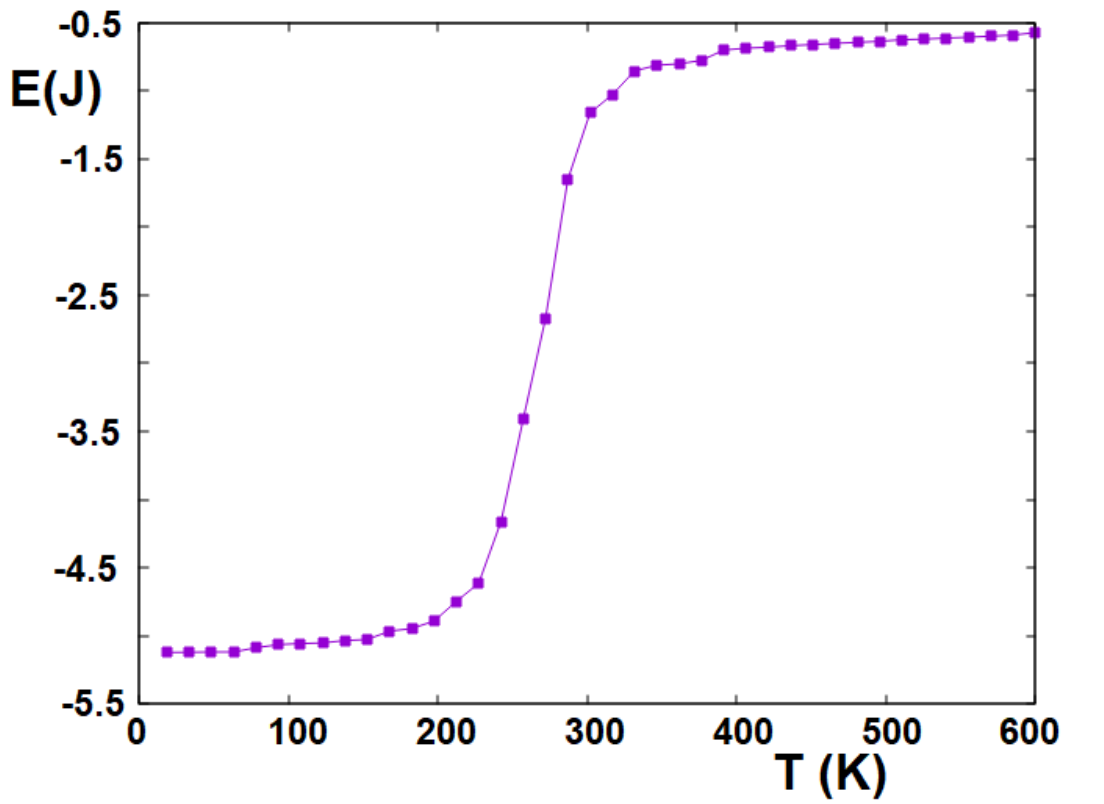}
\includegraphics[width=8cm,angle=0]{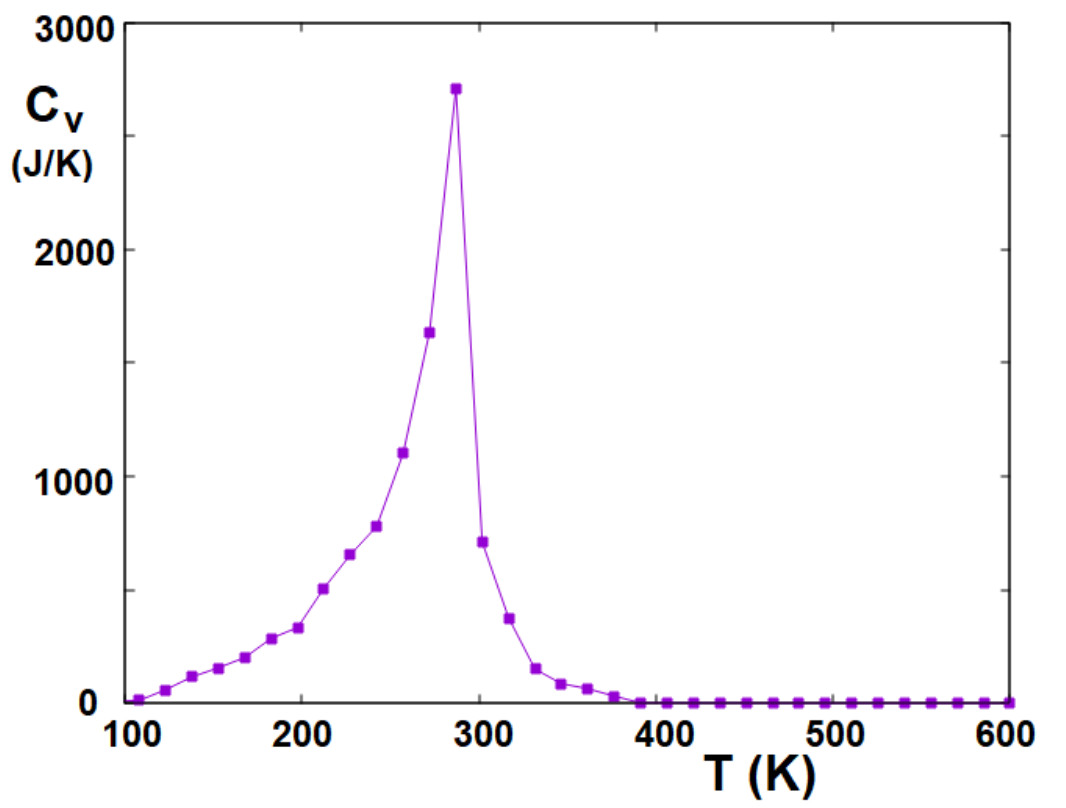}
\caption{(a) Internal energy per spin $E$ (in unit of $J$) versus  temperature $T$ in Kelvin, see real unit of $E(T=0)$ in Eq. (\ref{MCE}), (b) Specific heat  (in unit of J/Kelvin) versus $T$ calculated by energy fluctuations. Line is guide to the eye.}
\label{fig5}
\end{figure}

We show in Fig. \ref{fig5} the internal energy $E$ in unit of $J$ versus $T$ in Kelvin and the specific heat $C_V(T)$ in unit of $J/Kelvin$.  We will calculate $E$ in real unit later. As seen in this figure, the energy is continuous at $T_C$, so that the first-order transition at $T_C$ is excluded. Note that  $C_V(T)$ has a a very high peak. As said earlier, due to the strong fluctuations induced by the multi-spin interaction, the system energy highly fluctuates giving rise to a strong peak of $C_V$ which is calculated by using the energy fluctuations (Eq.  \ref{CV}). If $C_V$ is calculated by the derivative of the "average" $E$ with respect to $T$, the fluctuations are erased away and we do not have such a huge peak of $C_V$.   

\section{Discussion}\label{discuss}

\subsection{How to calculate the exchange interactions in real unit}

Experiments found that $J_2$ (interaction between Mn$^{3+}$ and Mn$^{4+}$   dominate and give rise to the ferromagnetic ordering up to very high temperatures $T_C=291$ K for  Pr$_{0.55}$Sr$_{0.45}$Mn$_{0.55}^{3+}$Mn$_{0.45}^{4+}$O$_{3}$. 

Since the ordering is ferromagnetic in spite of the fact that the interactions in the system have diiferent signs (FM and AF), we can consider the system as an "effective" ferromagnet.  We can  estimate the amplitudes of  physical parameters in real units using the following formula of the mean-field approximation for ferromagnets [13,26]:
\begin{equation}\label{MFTC}
T_C=\frac{2}{3k_B}ZS_{eff}(S_{eff}+1)J_{eff}
\end{equation}
where $Z=6$Mn+$8\times 0.55$Pr=10.4  is the effective coordination number at a Mn  site  and $S_{eff}$  the effective spin length which can be taken as the average on the Mn$^{4+}$, Mn$^{3+}$ and Pr$^{3+}$ using their concentrations: $S_{eff}=(0.55\times2+0.45\times1.5+0.55\times 1)/(0.55+0.45+0.55)=1.5$. Putting $T_C=291$ K in Eq. (\ref{MFTC}), we obtain $J_{eff}\simeq =0.000964$ eV$\simeq 9.64$ K.  In magnetic materials with Curie temperatures at room or higher temperatures the exchange interaction is of the order of several dozens of Kelvin [26,27]. The value found here is of the same order.  Note that  $J_{eff}$ is the effective exchange which is an average of various interactions by the mean-field theory  outlined above. The  interactions $J_1, ...,J_6$  can be caculated by using the value of $J$. For example, the dominant one is $J_2=0.54J=0.54\times291/0.7832\simeq 200 $ K, and the weak AF interaction between Mn ions of the same kind is $J_1=J_3=-0.1 J= -0.1\times 291/0.7832\simeq 37$ K.  
We however emphasize that using such a mean-field approximation, we obtain  the order of magnitude of interaction parameters, but not to a good precision since the mean-field approximation neglects fluctuations which are very strong in the present system (see the validity of the mean-field approximation, chapter 4 "Mean-Field Approximation" of [13]).

The value of the energy in the real unit (eV) can be estimated as follows: using  the relation between the classical ground-state  energy $E_0$ and the mean-field value of $J_{eff}$ calculated above, we have the value of $E_0$  given by

\begin{eqnarray}
E_0(MF)&=&-0.5\times Z\times S_{eff}^2 \times  J_{eff}\\
&=&-0.5\times 10.4\times1.5^2 \times 0.000964\times10^{-5}\  \mbox{eV}\\
&=&-0.01128\times10^{-5} \mbox{ eV}\\
&\simeq&-11.28\  \mbox{meV}\label{MCE}
\end{eqnarray}

%



\subsection   {Magnetocaloric Effect - Magnetic Entropy Change}

\begin{figure}[ht]
\centering
\includegraphics[width=6.6cm,angle=0]{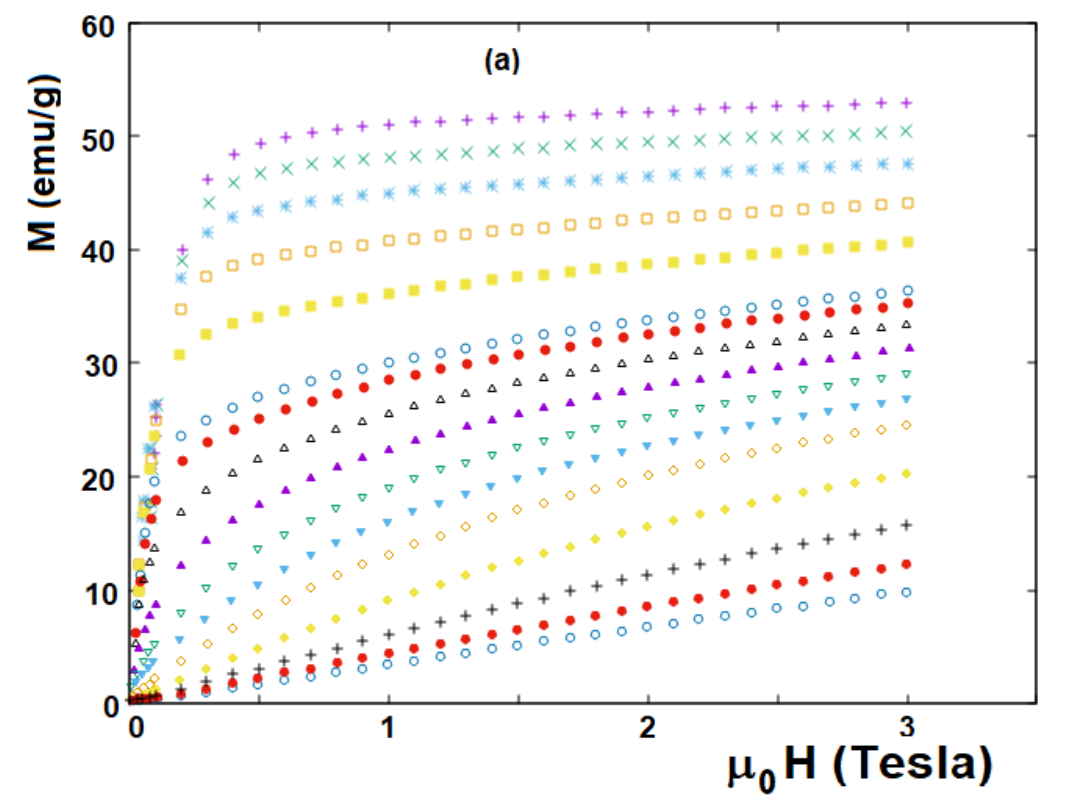}
\includegraphics[width=6.6cm,angle=0]{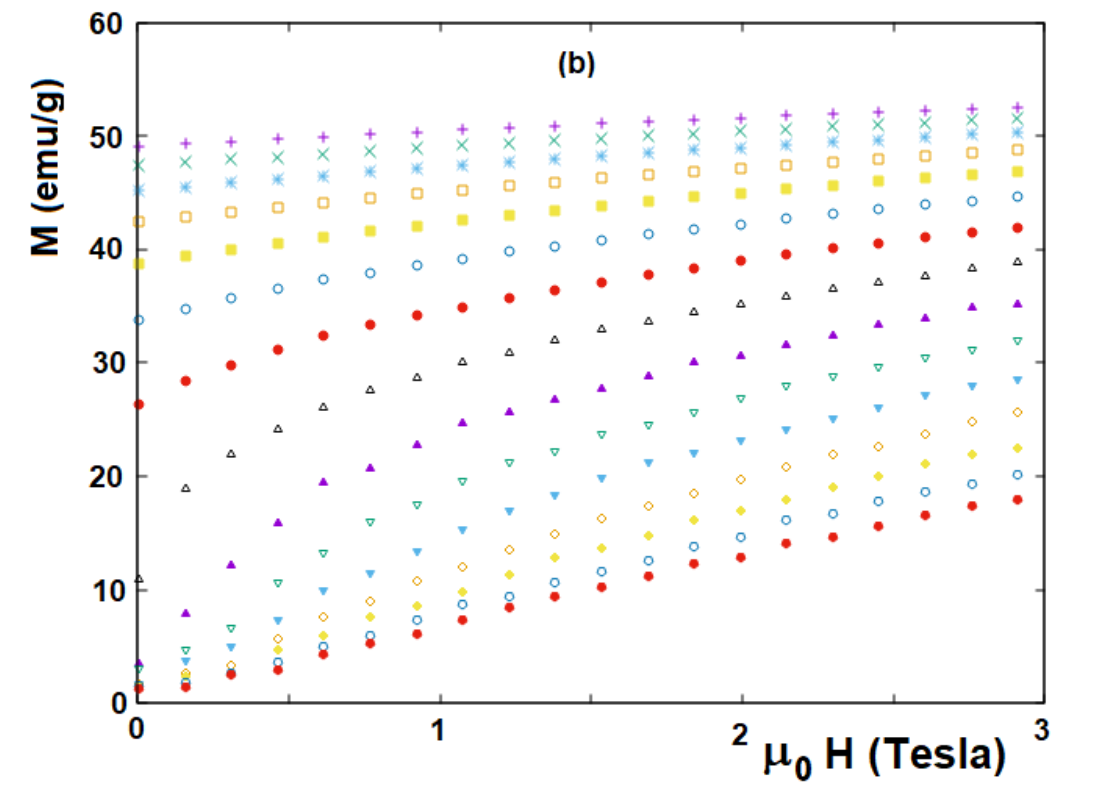}
\caption{ (a) Experimental magnetization for  $\mu_0H=0.03, 1, 2, 3$ Tesla at temperatures  $T$=230 K, 240  K, 250 K, 260 K,270 K,280 K, 282 K, 286 K, 290 K, 294 K, 298 K 302 K, 310 K, 320 K, 330 K, 340 K (from top to bottom). Results extracted from Fig. 1 of  Ref. [10], (b) MC results of magnetization versus $\mu_0H$ for $T$ ranging from 230 K, 239.5 K,...,363 K (from top to bottom, every 9.5 K).}
\label{fig6}
\end{figure}

Let us investigate the magneto-caloric effect by calculating the magnetization as a function of the magnetic field at several given temperatures ranging from the ferromagnetic phase to the paramagnetic phase, across the transition temperature. 
We show in Fig. \ref{fig6} the experimental magnetization versus the magnetic field $\mu_0H$ ranging from 0.03 to 3 Tesla, for $T$ ranging from 230 K to 340 K (from top curve to bottom one with a step of 10K). The MC results of the magnetization are also shown for comparison, with $T$ ranging from 230 K to 363 K with a step of 9.5 K (note that the value of the temperature step is not important).

 We obtain a qualitative agreement between the two sets of curves.  We have some remarks:

i) The MC magnetization curves far below $T_C$ are larger than the  experimental curves at low $H$. We think that this is because experimental samples are polycrystalline, there are certainly domains, impurities, dislocations, ...resulting in low $M$ at low $H$ and low $T$, in contrast to  MC samples which are on a lattice,

ii) For $T$ close to $T_C$ and above $T_C$, the agreement between experiments and MC results is better.  

These  differences will cause a small difference in  the magnetic entropy change   $|\Delta S_m|$ which is shown in the  following.


The magnetic entropy change $|\Delta S_m|$ is  calculated by applying a field progressively from 0 to $ \mu_0 H$ at two given close temperatures $T_1$ and $T_2$. $|\Delta S_m|$ is given by the  thermodynamic Maxwell formula

\begin{equation}\label{DeltaS}
|\Delta S_m (T,H)|=\int_0^H\left[\frac{\delta M(T,H_i)}{\delta T} \right]_{H_i}\  \mu_0 dH_i
\end{equation}
where $\delta M$ is the change of the magnetization at $H$ when $T$ varies from $T_1$ to $T_2$, namely $T\rightarrow T+\delta T$.  This formula is discretized  and used in experiments as well as in MC simulations as

\begin{equation}\label{DeltaS1}
|\Delta S_m (T,H)|=\sum_i\left[\frac{M_i-M_{i+1}}{T_{i+1}-T_i} \right]\mu_0 \Delta H_i
\end{equation}

\begin{figure}[ht]
\centering.
\includegraphics[width=7cm,angle=0]{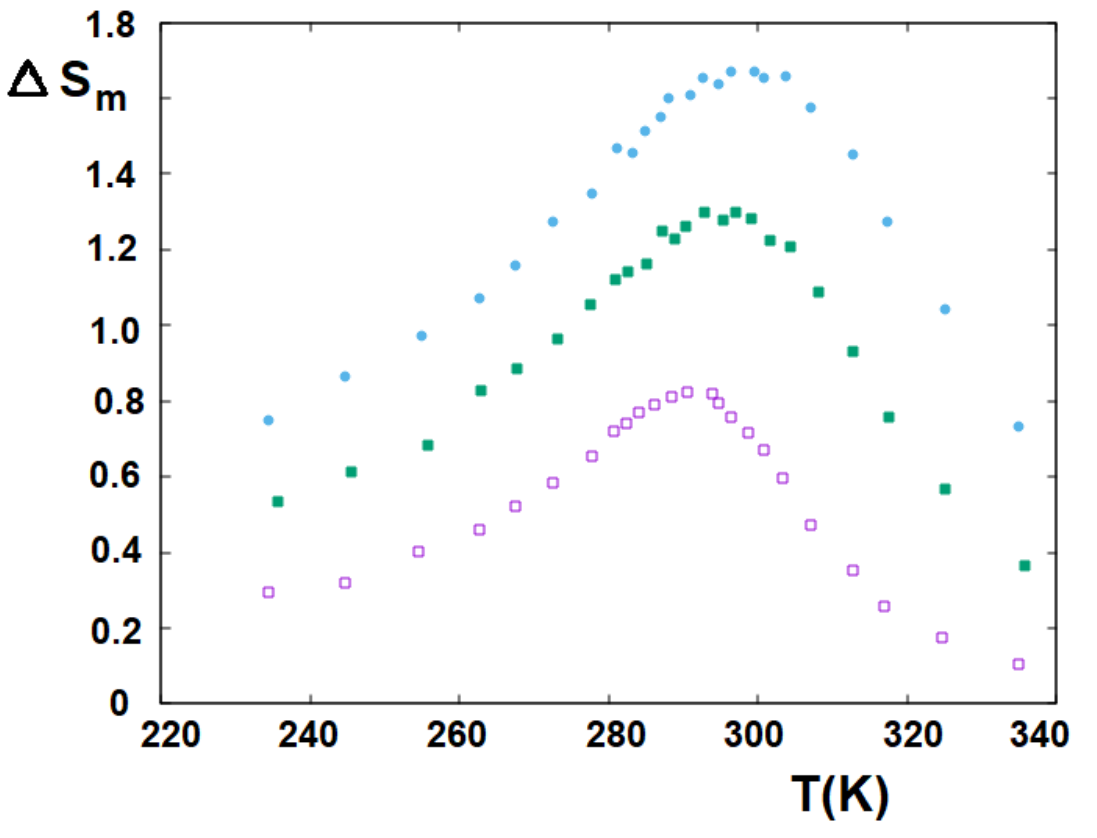}
\caption{ Experimental magnetic entropy change $|\Delta S_m|$ (J/(kgKelvin) versus  temperature $T$ for $\mu_0H=$1, 2, 3 Tesla (from bottom to top), extracted from Fig. 2b of Ref. [10]. See text for comment.}
\label{figDeltaS}
\end{figure}

The experimental magnetic entropy change $|\Delta S_m|$ using this formula   is shown in Fig. \ref{figDeltaS} for $\mu_0H$=1, 2, 3 Tesla: the peak temperature increases slightly with increasing $H$, passing from $T=291.5$K for $\mu_0H=1$ Tesla to $T=299.5$ K for $mu_0H=3$ Tesla. Note that the peak height increases with increasing $H$. 

\begin{figure}[ht]
\centering.
\includegraphics[width=7cm,angle=0]{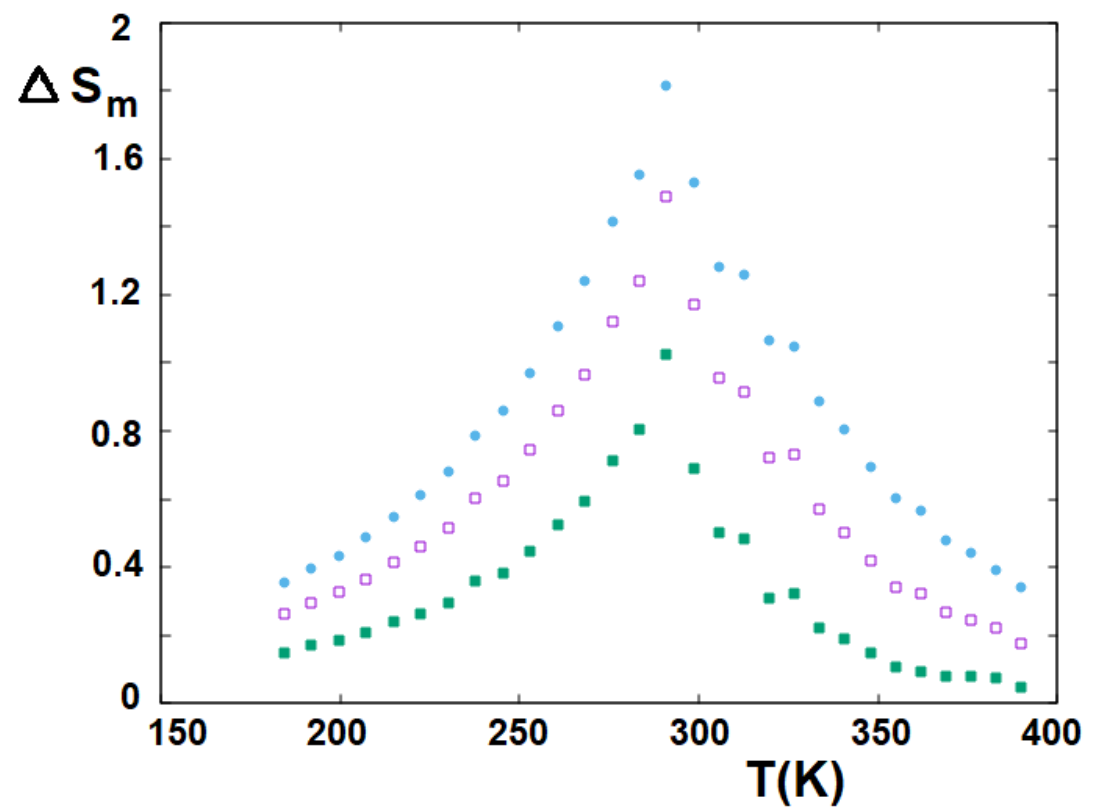}
\caption{ MC result for magnetic entropy change $|\Delta S_m|$ (J/(kgKelvin) versus  temperature $T$ for $\mu_0H=$1, 2, 3 Tesla (from bottom to top). See text for comment.}
\label{figMCDeltaS}
\end{figure}

To compare with experimental magnetic entropy change, we carry out the calculation of  $|\Delta S_m|$ using Eq. (\ref{DeltaS1}) as in experiments with the MC magnetizations as shown in Fig. \ref{fig6}a, but with a smaller temperature step, i. e. 4.75 K, for a better precision (we did not show the whole set of curves in  Fig. \ref{fig6}a for clarity of the figure).  This gives the result shown in Fig. \ref{figMCDeltaS}.  Several remarks are in order:

i) $|\Delta S_m|$ obtained from MC simulations show a peak at each value of $\mu_0H$ from 1 Tesla to 3 Tesla. The value of the peak increases with increasing $H$, in agreement with experiments shown in Fig. \ref{figDeltaS}. The values of the peak height are very close to those of experiments. 

ii) The peak temperarure does not however change  with increasing $H$ as observed in experiments. To explain this difference, we have looked at the experimental results in the paper [28] performed on the same compound with a different concentration Pr$_{0.67}$Sr$_{0.33}$Mn$_{0.67}^{3+}$Mn$_{0.33}^{4+}$O$_{3}$. In [28] several samples have been realized: as one goes from samples with impuriries, defects,... to the pure one (sample P1200), the peak temperatures do not change significantly with $H$ for the pure sample and the peak widths become then less wide (see Fig.  9 of Ref. [28]). This is what we find with MC samples which are exempt of defects, dislocations, ...  In view of experimental data obtained for many samples of  Pr$_{0.67}$Sr$_{0.33}$Mn$_{0.67}^{3+}$Mn$_{0.33}^{4+}$O$_{3}$ published in Ref. [28], we believe that the results published in Ref. [11] for  Pr$_{0.55}$Sr$_{0.45}$Mn$_{0.55}^{3+}$Mn$_{0.45}^{4+}$O$_{3}$ are from a sample with more domains, defects, ... than the sample used in Ref. [10] which results in data in agreement with our MC simulations.

The formula used to calculate the Relatice Cooling Power (RCP) is 

\begin{equation}\label{RCP}
RCP(H)=|\Delta S_{max} (H)|\times \Delta T
\end{equation}
where $|\Delta S_{max} (H)|$ is the maximum value of $|\Delta S_{m} (H)|$ and  $\Delta T$  the temperature range at  the full
width at half maximum.  

We show in Table \ref{figRCP} the  experimental Relative Cooling Power (RCP) extracted from Fig. 4 of Ref. [10] and the MC RCP calculated from the curves of Fig. \ref{figMCDeltaS}. As seen, there is a slight difference between experiments and MC results. This difference comes from slightly higher peak values in MC data (Fig. \ref{figMCDeltaS}). We believe that this is due to the fact that MC samples are exempt of discolations, impurities, ... unlike experimental samples. 

The compound we studied here has  high RCP which is very important for high efficiency in applications using the magnetocaloric effect.


\begin{table}
\centering
\caption{Experimental Relative Cooling Power and MC results of Pr$_{0.55}$Sr$_{0.45}$Mn$_{0.55}^{3+}$Mn$_{0.45}^{4+}$O$_{3}$, for  $\mu_0H$=1, 2, 3 Tesla. See text for comments.}
\label{figRCP}       
\begin{tabular}{lll}
\\
\hline
\\
$\mu_0H$(T) |& Experimental RCP(J/kg) |& MC RCP (J/kg)  \\
\\
\hline\\
1 & 44.1 & 54.6 \\
2 & 84.5 & 105.7 \\
3& 143.6 & 163.8 \\
\\\hline
\end{tabular}
\end{table}

To conclude this section, let us mention several works using a different approach: in Ref. [29 ] the authors used first the DFT calculations to estimate the pairwise exchange interactions between magnetic ions and use them in MC simulations to study magnetic, magnetocaloric and thermoelectric properties of perovskite LaFeO3 compound. They found the magnetization at $T=0$ and the value of the Neel tremperature in agreement with experiments. There is however no information on how the experimental magnetization depends on $T$. In Refs. [30-31] the same approach has been applied to   study properties of spinel ferrites. Again in these cases, there is no curve of experimental $M(T)$ which allows for a precise modeling of  the detailed interactions in the systems. Our strategy, as shown above, is the reverse: we find first a model Hamiltonian which reproduces the experimental $M(T)$. Then, by fitting with experiments we find the values of all  interactions in the model.

\section{Conclusion}\label{concl}

We have compared in this paper our MC results with the magnetic properties of the perovskite compound Pr$_{0.55}$Sr$_{0.45}$Mn$_{0.55}^{3+}$Mn$_{0.45}^{4+}$O$_{3}$ observed experimentally [10]. In order to get an agreement with the experimental magnetization over the whole temperature range, we have included a new multi-spin interaction in the Hamiltonian and we have used the multi-state Ising model. The experimental transition is likely of first order in view of the vertical fall of the magnetization at $T_C$. However, MC results on other physical quantities such as the internal energy and the sublattice spins do  show  discontinuities at $T_C$ in spite of the fact that  the specific heat and the susceptibility show very high peaks (but with a relatively large width).  The origin of the huge peaks has been discussed in the light of spin fluctuations caused by the multi-spin interaction.  Since the aim of this paper was not to determine for sure the nature of the phase transition by advanced MC techniques such as the histogram method [23,24] or the Wang-Landau algorithm [25], we leave this point for a future study.

We have studied the magneto-caloric effect by calculating the magnetization as a function of the magnetic field $\mu_0H$  for several temperatures across the phase transition.  From these curves, we calculated  the magnetic entropy change $|\Delta S_{m} (H)|$  as a function of the magnetic field, and  we deduced the RCP. We have found a qualitative agreement with experimental data. The difference is very small. The model used here is therefore justified in view of the consistence with experiments.  
We emphasize that the main ingredient in the model is the multi-spin interaction. It allows to reproduce the experimental magnetization below $T_C$ and other experimental data. For a system of high disorder such as the present system, without the multi-spin interaction, we cannot find such a sharp phase transition.

\vspace{2cm}

\acknowledgments

Yethreb Essouda  is indebted to the CY Cergy Paris University for hospitality during her working visits.\\

{ \bf Authors' statement}: There is no conflict of interest concerning this work. Y. Essouda analyzed experimental and numerical data. H. T. Diep conceived the modeling, wrote the MC program and the first version of the paper. M. Ellouze analyzed experimental data and corrected and re-edited the manuscript.

\end{document}